\documentclass[english,letterpaper,english,prd,preprintnumbers,nofootinbib,showpacs,shortbibliography,notitlepage]{revtex4-1}
\usepackage[T1]{fontenc}
\usepackage[latin9]{inputenc}
\usepackage{verbatim}
\usepackage{amsmath}
\usepackage{amssymb}
\usepackage{graphicx}
\usepackage{babel}
\begin{document}
\global\long\def\edge#1{\left.#1\right|}

\global\long\def\d{\mathrm{d}}
\global\long\def\yi{\varphi}

\global\long\def\bra#1{\left\langle #1\right|}
\global\long\def\ket#1{\left| #1 \right\rangle }

\global\long\def\tr{\mbox{Tr}}

\preprint{Alberta Thy 28-16}

\title{Decay of the dimuonium into a photon and a neutral pion}

\author{Andrzej Czarnecki }

\affiliation{Department of Physics, University of Alberta, Edmonton, Alberta,
Canada T6G 2E1}

\author{Savely~G.~Karshenboim}
\email{savely.karshenboim@mpq.mpg.de}

\selectlanguage{english}%

\affiliation{Ludwig-Maximilians-Universität, Fakultät für Physik, 80799 München,
Germany}

\affiliation{Pulkovo Observatory, St.~Petersburg, 196140, Russia}

\affiliation{Max-Planck-Institut für Quantenoptik, 85748 Garching, Germany}
\begin{abstract}
We compute the decay rate of dimuonium into a neutral pion and a photon.
We find that approximately one in $10^{5}$ ortho-dimuonia decays
into this channel. We also determine the contribution of the virtual
photon-pion loop to the hyperfine splitting in dimuonium and reproduce
its leading effect in the anomalous magnetic moment of the muon.
\end{abstract}
\maketitle

\section{Introduction}

Dimuonium Dm (also known as true muonium) is a bound state of a muon
and an antimuon, analogous to positronium but about 207 times heavier
\cite{Bilenky:1969zd,hughes:1971aa}. While positronium was discovered
already 65 years ago \cite{Deutsch:1951zza}, dimuonium has not been
observed yet. Recently, however, the prospect for its discovery has
become brighter: it may be produced in experiments searching for exotic
light bosons \cite{Banburski:2012tk}. Production of Dm at rest is
under consideration at the $e^{+}e^{-}$ collider at the Budker Institute of Nuclear Physics in Novosibirsk
\cite{simonPriv}. On the theory side, Dm production \cite{Sakimoto:2015vrf,Brodsky:2009gx,Ginzburg:1998df,Nemenov:1972ph},
spectrum \cite{Ji:2016fat,Lamm:2015fia,Jentschura:1997tv,Karshenboim:1998we,karshenboim:1998aa},
and decays \cite{Karshenboim:1998we,Ginzburg:1998df,Jentschura:1997tv}
have been studied. The name ``dimuonium'' was first introduced in
\cite{Malenfant:1987tm}.

\begin{figure}[h]
\centering

\includegraphics[scale=0.35]{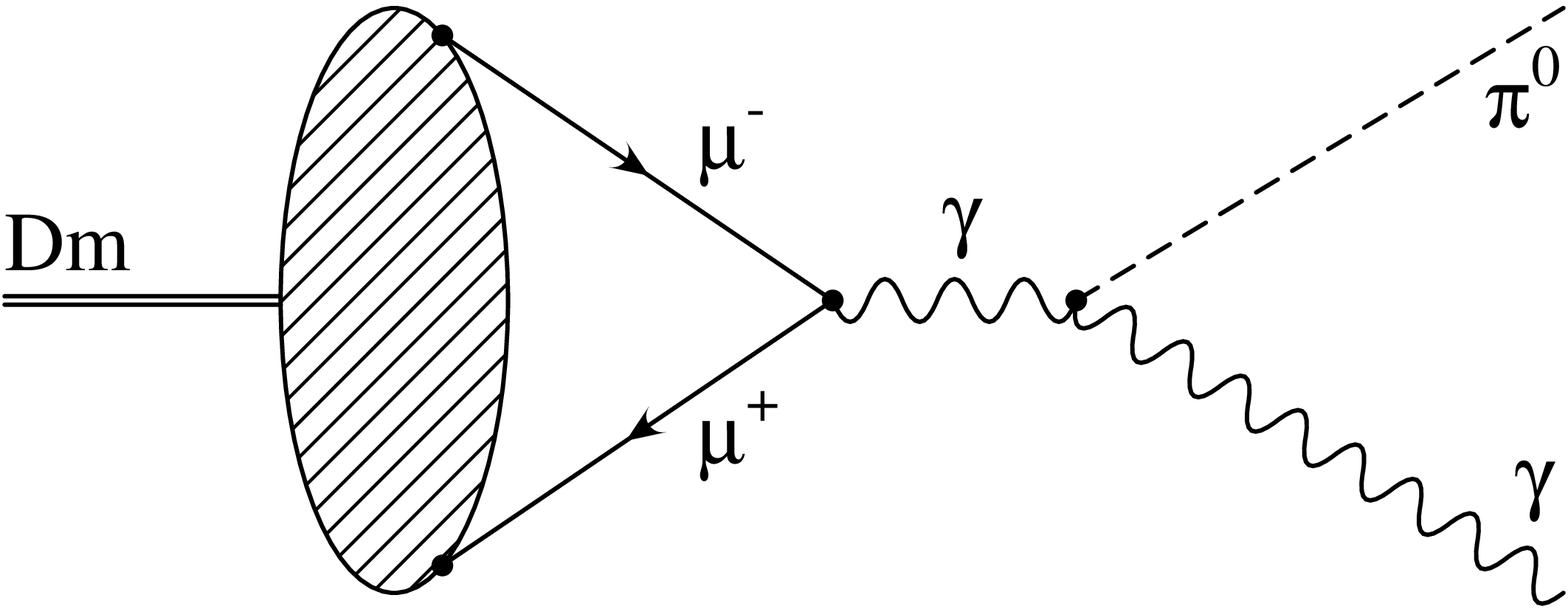}\hspace*{5mm}\includegraphics[scale=0.35]{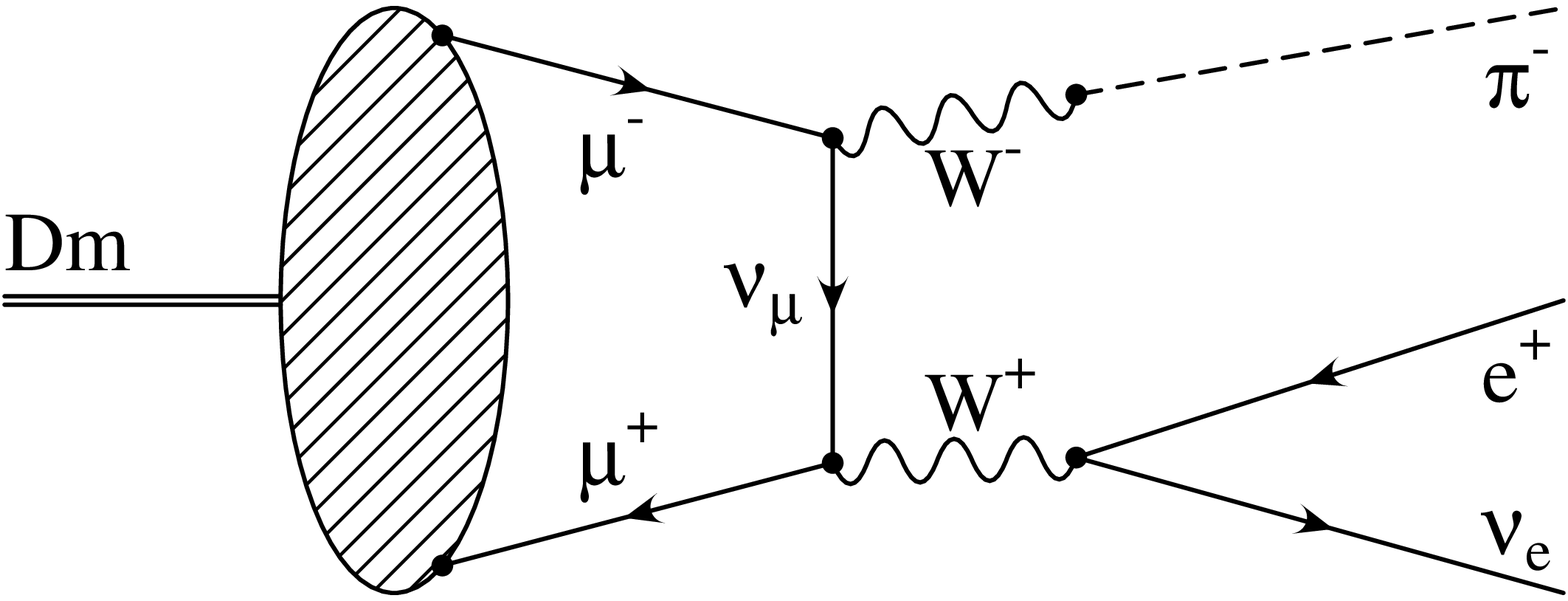}

\hspace*{10mm}(a)\hspace*{80mm}(b)

\caption{Decays of dimuonium with pions in the final state: (a) Dm$\to\pi^{0}\gamma$
considered in this work; (b) Decay with a charged pion. It is strongly
suppressed because of the large mass of the $W$ boson and neglected
here.\label{fig:Decays-of-dimuonium}}

\end{figure}

Although Dm is a purely leptonic system, it is affected by hadrons
through higher-order effects. It is the lightest pure leptonic system
with hadronic decay channels. Here we present the rate of the dimuonium
decay into a neutral pion $\pi^{0}$ and a photon $\gamma$, shown
in Fig.~\ref{fig:Decays-of-dimuonium}(a), a decay channel that
has not been considered so far. It is interesting because it is a
new two-body decay of the spin-triplet dimuonium (ortho-dimuonium,
o-Dm), with a clean signature: a monochromatic photon. Such hadronic
final states are not accessible to positronium because of its small
mass. There is also another hadronic decay channel, with a charged pion,
shown in Fig.~\ref{fig:Decays-of-dimuonium}(b);
and an analogous channel with the opposite-sign pion. However,
these processes are additionally suppressed by inverse powers of the $W$
boson mass and are extremely rare.

\section{Decay rate of $\text{Dm}\to\pi^{0}\gamma$}

We use the amplitude of the $\pi^{0}\gamma\gamma$ coupling (throughout
this paper we use $\hbar=c=1$),
\begin{equation}
\mathcal{M}=\frac{\alpha F_{\pi^{0}\gamma\gamma}\left(q_{1}^{2},q_{2}^{2}\right)}{\pi F_{\pi}}\epsilon^{\mu\nu\alpha\beta}\epsilon_{1\mu}^{\star}\epsilon_{2\nu}^{\star}q_{1\alpha}q_{2\beta},
\end{equation}
where $\alpha\simeq1/137$ is the fine structure constant, $F_{\pi}\simeq92$
MeV is the pion decay constant, and $F_{\pi^{0}\gamma\gamma}\left(q_{1}^{2},q_{2}^{2}\right)$
is a form factor that we need only with one of the photons on-shell;
we use a simple vector meson dominance approach to model this form
factor,
\begin{equation}
F_{\pi^{0}\gamma\gamma}\left(q^{2},0\right)\simeq\frac{1}{1-\frac{q^{2}}{M^{2}}},
\end{equation}
where the mass $M$ is on the order of the $\rho$-meson mass, $M\simeq m_{\rho}\simeq769$
MeV. We find the width
\begin{equation}
\Gamma(\text{o-Dm}\to\pi^{0}\gamma)=\frac{\alpha^{6}E^{3}}{48\pi^{3}F_{\pi}^{2}}\left(\frac{1}{1-\frac{4m_{\mu}^{2}}{m_{\rho}^{2}}}\right)^{2},\label{GammaDm}
\end{equation}
where o-Dm refers to ortho-dimuonium, $m_{\mu}$ is the muon mass,
and $E$ is the energy of the photon in the final state,
\begin{equation}
E\simeq\frac{4m_{\mu}^{2}-m_{\pi}^{2}}{4m_{\mu}}\simeq63\text{ MeV}.
\end{equation}
The dominant decay channel of ortho-dimuonium is into an $e^{+}e^{-}$
pair, with the rate
\begin{equation}
\Gamma\left(\text{o-Dm}\to e^{+}e^{-}\right)=\frac{\alpha^{5}m_{\mu}}{6}.
\end{equation}
The ratio of the decay rates into $\pi^{0}\gamma$ and the $e^{+}e^{-}$
is
\begin{equation}
\frac{\Gamma\left(\text{o-Dm}\to\pi^{0}\gamma\right)}{\Gamma\left(\text{o-Dm}\to e^{+}e^{-}\right)}=\frac{\alpha\left(4m_{\mu}^{2}-m_{\pi}^{2}\right)^{3}}{512\pi^{3}F_{\pi}^{2}m_{\mu}^{4}}\left(\frac{1}{1-\frac{4m_{\mu}^{2}}{m_{\rho}^{2}}}\right)^{2}\simeq0.9\cdot10^{-5}.
\end{equation}
This branching ratio is small because the coupling of the neutral
pion to photons can be interpreted as a quantum (loop-induced) effect
\cite{Steinberger:1949wx}. There are additional suppression factors
such as the reduced phase space volume. However, there does not seem
to be another two-body decay channel accessible to ortho-dimuonium
and not to the spin singlet (para-dimuonium), p-Dm (the dominant final
state for o-Dm, $e^{+}e^{-}$, can be reached from p-Dm via a two-photon
annihilation). Thus in principle the peak of 63 MeV photons (in the
rest-frame of the decaying o-Dm) can be used to establish the presence
of ortho-dimuonium.

The decay rate into $\pi^{0}\gamma$ is also much smaller than into
three photons,
\begin{equation}
\frac{\Gamma(\text{Dm}\to\pi^{0}\gamma)}{\Gamma(\text{Dm}\to\gamma\gamma\gamma)}
=\frac{3
  \left(1-\frac{m_{\pi}^{2}}{4m_{\mu}^{2}}\right)^{3} m_{\mu}^{2}
           }
{32\pi^{2}\left(\pi^{2}-9\right)
  \left(1-\frac{4m_{\mu}^{2}}{m_\rho^{2}}\right)^2 F_{\pi}^{2} }
\simeq 0.3\%.
\end{equation}

\section{Contribution to the hyperfine splitting}

\begin{figure}[h]
\centering

\includegraphics[scale=0.4]{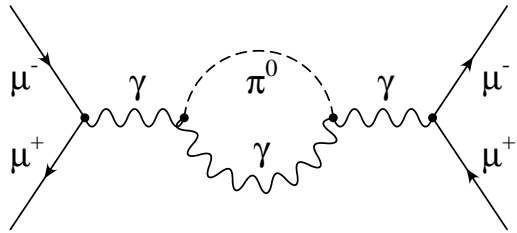}

\caption{Virtual annihilation into $\pi^{0}\gamma$ is possible only for ortho-dimuonium
and thus contributes to the hyperfine splitting of Dm.\label{fig:Virtual-annihilation-into}}
\end{figure}

Since the virtual one-photon annihilation is possible only for ortho-dimuonium,
the  $\pi^{0}\gamma$ loop contributes to the hyperfine splitting (HFS)
through the diagram shown in Fig.~\ref{fig:Virtual-annihilation-into}.
This is an additional hadronic vacuum polarization contribution to
the HFS, not considered in previous calculations, such as for example
\cite{Jentschura:1997tv}.

We find this contribution to slightly modify the one-photon virtual
annihilation,
\begin{equation}
\Delta E=\frac{\alpha^{4}m_{\mu}}{4}\left\{ 1+\frac{\alpha^{2}m_{\rho}^{2}}{32\pi^{4}F_{\pi}^{2}}\left[\frac{2\int_{0}^{1}x\left[\epsilon-\left(1-x\right)\delta\right]\ln\left|1+\frac{1-x}{x\left[\epsilon-\left(1-x\right)\delta\right]}\right|\text{d}x-1-\frac{i\pi}{3}\left(1-\frac{\epsilon}{\delta}\right)^{3}\delta}{\left(1-\delta\right)^{2}}+\frac{1-\epsilon+\epsilon\ln\epsilon}{\left(1-\epsilon\right)^{2}}\right]\right\}
\end{equation}
where $\delta=\frac{4m_{\mu}^{2}}{m_{\rho}^{2}}\simeq0.08$ and $\epsilon=\frac{m_{\pi}^{2}}{m_{\rho}^{2}}\simeq0.03$.
The imaginary part reproduces the decay rate, $\Gamma(\text{o-Dm}\to\pi^{0}\gamma)=-2\text{Im}\Delta E$,
in agreement with Eq.~\eqref{GammaDm}. The real part gives a correction
to the HFS; it is well approximated by dropping $x\left(1-x\right)\delta$
in the argument of the log and retaining only terms of first order
in $\epsilon$ and $\delta$ in the remaining result,
\begin{align}
\text{Re}\Delta E & \simeq\frac{\alpha^{4}m_{\mu}}{4}\left[1-\frac{\alpha^{2}m_{\rho}^{2}}{32\pi^{4}F_{\pi}^{2}}\left(\epsilon\ln\frac{1}{\epsilon}+\frac{\epsilon}{2}+2\delta\right)\right]\\
 & \simeq\frac{\alpha^{4}m_{\mu}}{4}\left[1-3.\cdot10^{-7}\right]
\end{align}
The numerical value of this 0.3 part per million shift is about $-6$
MHz.

\section{Contribution to the muon anomalous magnetic moment}

\begin{figure}[h]
\centering

\includegraphics[scale=0.4]{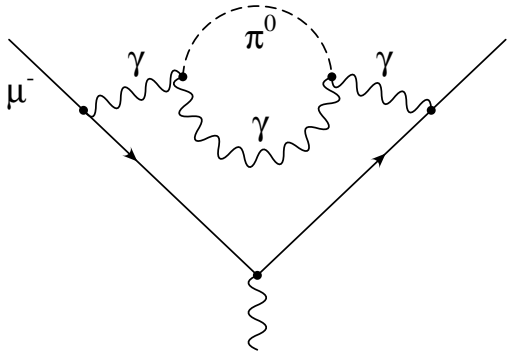}

\caption{$\pi^{0}\gamma$ contribution to the muon anomalous magnetic moment.\label{fig:MagneticAnomaly}}
\end{figure}

The $\pi^{0}\gamma$ loop contribution to the vacuum polarization
modifies also the anomalous magnetic moment of the muon, $a_{\mu}=\frac{g_{\mu}-2}{2}$.
This is realized by closing the antimuon line in Fig.~\ref{fig:Virtual-annihilation-into}
and connecting it to an external magnetic field, as shown in Fig.~\ref{fig:MagneticAnomaly}.
This effect was determined in \cite{Blokland:2001pb} using an expansion
in two small parameters, the mass ratio $m_{\mu}/m_{\rho}$ and the
normalized difference of the pion and the muon masses squared, $\frac{m_{\pi}^{2}-m_{\mu}^{2}}{m_{\mu}^{2}}$
(see also \cite{Achasov:2002bh,Achasov:2002xm,Marciano:2016yhf,Dorokhov:2012qa}).
We can now check that result numerically. To this end, we consider
the $\pi^{0}\gamma$ contribution to the polarization,
\begin{equation}
\Pi_{R}^{\mu\nu}\left(p\right)=i\left(p^{2}g^{\mu\nu}-p^{\mu}p^{\nu}\right)\Pi_{R}\left(p^{2}\right)
\end{equation}
where, with the on-shell charge renormalization and to the leading
order in $1/M^{2}$ (we use $M$ again for the vector meson mass in the form factor,
to emphasize its role as the ultraviolet cutoff),
\begin{align}
\Pi_{R}\left(p^{2}\right) & =\frac{\alpha^{2}}{16\pi^{4}F_{\pi}^{2}}\int_{0}^{1}\left\{ \frac{1}{\left(1-\frac{p^{2}}{M^{2}}\right)^{2}}\left[A\left(0,p^{2}\right)\ln\frac{A\left(M^{2},p^{2}\right)}{A\left(0,p^{2}\right)}-\frac{M^{2}}{2}\right]+\frac{M^{2}}{2}-A\left(0,0\right)\ln\frac{A\left(M^{2},0\right)}{A\left(0,0\right)}\right\} \d x\\
 & \equiv\frac{\alpha^{2}}{16\pi^{4}F_{\pi}^{2}}p^{2}J\left(p^{2}\right)\\
A\left(M^{2},p^{2}\right) & \equiv xm_{\pi}^{2}+\left(1-x\right)M^{2}-x\left(1-x\right)p^{2}-i0.
\end{align}
It is useful to isolate $p^{2}$
as an explicit overall factor by rewriting $\Pi_{R}$. In the leading order
in $1/M^{2}$, the integral in $\Pi_{R}$ is
\begin{equation}
p^{2}J\left(p^{2}\right)=\int_{0}^{1}\left[-p^{2}-x\left(1-x\right)p^{2}\ln\frac{A\left(M^{2},p^{2}\right)}{A\left(0,p^{2}\right)}+A\left(0,0\right)\ln\frac{A\left(M^{2},p^{2}\right)}{A\left(0,p^{2}\right)}\frac{A\left(0,0\right)}{A\left(M^{2},0\right)}\right]\d x.
\end{equation}
Integrating by parts, we find
\begin{align}
\int_{0}^{1}\left[-x\left(1-x\right)p^{2}\ln\frac{A\left(M^{2},p^{2}\right)}{A\left(0,p^{2}\right)}+A\left(0,0\right)\ln\frac{A\left(M^{2},p^{2}\right)}{A\left(M^{2},0\right)}\frac{A\left(0,0\right)}{A\left(0,p^{2}\right)}\right]\d x\qquad\qquad\qquad\qquad\qquad\qquad\\
=\frac{p^{2}}{6}\int_{0}^{1}\left[\frac{\left[\left(1-x\right)^{2}M^{2}-x^{2}m_{\pi}^{2}\right]\left[\left(3-2x\right)\left(1-x\right)M^{2}+x^{2}m_{\pi}^{2}\right]}{\left[\left(1-x\right)M^{2}+xm_{\pi}^{2}\right]\left(p^{2}-\frac{m_{\pi}^{2}}{1-x}-\frac{M^{2}}{x}\right)}+\frac{m_{\pi}^{2}x^{3}}{p^{2}-\frac{m_{\pi}^{2}}{1-x}}\right]\frac{\d x}{\left(1-x\right)^{2}}.
\end{align}
We recognize in the $p^{2}$-dependent denominators a similarity to
propagators of massive particles. A vector boson of mass $m$, with
a photon-like coupling to the muon, contributes to its anomalous magnetic
moment the amount \cite{Brodsky:1967sr}
\begin{equation}
\Delta a_{\mu}=\frac{\alpha}{\pi}\int_{0}^{1}\d y\frac{y^{2}\left(1-y\right)}{y^{2}+\left(1-y\right)\frac{m^{2}}{m_{\mu}^{2}}},
\end{equation}
so the effect of the $\pi^{0}\gamma$ loop is
\begin{align}
\Delta a_{\mu}\left(\pi^{0}\gamma\right) & =\frac{\alpha^{3}}{96\pi^{5}F_{\pi}^{2}}\left[2m_{\mu}^{2}+\int_{0}^{1}\frac{\d x}{\left(1-x\right)^{2}}\int_{0}^{1}\d yy^{2}\left(1-y\right)\right.\\
 & \qquad\left.\times\left\{ \frac{\left[\left(1-x\right)^{2}M^{2}-x^{2}m_{\pi}^{2}\right]\left[\left(3-2x\right)\left(1-x\right)M^{2}+x^{2}m_{\pi}^{2}\right]}{\left[\left(1-x\right)M^{2}+xm_{\pi}^{2}\right]\left[y^{2}+\left(1-y\right)\frac{xm_{\pi}^{2}+\left(1-x\right)M^{2}}{x\left(1-x\right)m_{\mu}^{2}}\right]}+\frac{m_{\pi}^{2}x^{3}}{y^{2}+\frac{1-y}{1-x}\frac{m_{\pi}^{2}}{m_{\mu}^{2}}}\right\} \right].
\end{align}
Numerically, in the limit $m_{\pi}=m_{\mu}$ and $M=m_{\rho}$, this
gives $\Delta a_{\mu}\left(\pi^{0}\gamma\right)=4.8\cdot10^{-11}$,
in agreement with eq.~(10) in \cite{Blokland:2001pb}. We have also
reproduced analytically the leading logarithm $\sim\ln\frac{M^{2}}{m_{\pi}^{2}}$
given in that work.

\section{Conclusions}

In eq.~\eqref{GammaDm} we have given the decay rate of a new decay
channel of orthodimuonium, $\text{o-Dm}\to\pi^{0}\gamma$. This channel
occurs approximately once every $10^{5}$ decays. As a byproduct of
this study, we have determined the correction of a virtual $\pi^{0}\gamma$
loop to the hyperfine splitting of dimuonium. Finally, as a check
of our results, we reproduced the leading contribution of this loop
to the anomalous magnetic moment of the muon and found agreement
with the original determination \cite{Blokland:2001pb}.

Acknowledgement: We thank A.~A.~Penin for helpful discussions. This
research was supported by Science and Engineering Research Canada
(NSERC), RFBR (under grant \# 15-02-00539), and DFG (under grant \#
KA 4645/1-1).


\begin{thebibliography}{21}%
\makeatletter
\providecommand \@ifxundefined [1]{%
 \@ifx{#1\undefined}
}%
\providecommand \@ifnum [1]{%
 \ifnum #1\expandafter \@firstoftwo
 \else \expandafter \@secondoftwo
 \fi
}%
\providecommand \@ifx [1]{%
 \ifx #1\expandafter \@firstoftwo
 \else \expandafter \@secondoftwo
 \fi
}%
\providecommand \natexlab [1]{#1}%
\providecommand \enquote  [1]{``#1''}%
\providecommand \bibnamefont  [1]{#1}%
\providecommand \bibfnamefont [1]{#1}%
\providecommand \citenamefont [1]{#1}%
\providecommand \href@noop [0]{\@secondoftwo}%
\providecommand \href [0]{\begingroup \@sanitize@url \@href}%
\providecommand \@href[1]{\@@startlink{#1}\@@href}%
\providecommand \@@href[1]{\endgroup#1\@@endlink}%
\providecommand \@sanitize@url [0]{\catcode `\\12\catcode `\$12\catcode
  `\&12\catcode `\#12\catcode `\^12\catcode `\_12\catcode `\%12\relax}%
\providecommand \@@startlink[1]{}%
\providecommand \@@endlink[0]{}%
\providecommand \url  [0]{\begingroup\@sanitize@url \@url }%
\providecommand \@url [1]{\endgroup\@href {#1}{\urlprefix }}%
\providecommand \urlprefix  [0]{URL }%
\providecommand \Eprint [0]{\href }%
\providecommand \doibase [0]{http://dx.doi.org/}%
\providecommand \selectlanguage [0]{\@gobble}%
\providecommand \bibinfo  [0]{\@secondoftwo}%
\providecommand \bibfield  [0]{\@secondoftwo}%
\providecommand \translation [1]{[#1]}%
\providecommand \BibitemOpen [0]{}%
\providecommand \bibitemStop [0]{}%
\providecommand \bibitemNoStop [0]{.\EOS\space}%
\providecommand \EOS [0]{\spacefactor3000\relax}%
\providecommand \BibitemShut  [1]{\csname bibitem#1\endcsname}%
\let\auto@bib@innerbib\@empty
\bibitem [{\citenamefont {Bilenky}\ \emph {et~al.}(1969)\citenamefont
  {Bilenky}, \citenamefont {Nguyen}, \citenamefont {Nemenov},\ and\
  \citenamefont {Tkebuchava}}]{Bilenky:1969zd}%
  \BibitemOpen
  \bibfield  {author} {\bibinfo {author} {\bibfnamefont {S.~M.}\ \bibnamefont
  {Bilenky}}, \bibinfo {author} {\bibfnamefont {V.~H.}\ \bibnamefont {Nguyen}},
  \bibinfo {author} {\bibfnamefont {L.~L.}\ \bibnamefont {Nemenov}},\ and\
  \bibinfo {author} {\bibfnamefont {F.~G.}\ \bibnamefont {Tkebuchava}},\
  }\href@noop {} {\bibfield  {journal} {\bibinfo  {journal} {Yad. Fiz.}\
  }\textbf {\bibinfo {volume} {10}},\ \bibinfo {pages} {812} (\bibinfo {year}
  {1969})},\ \bibinfo {note} {{Sov. J. Nucl. Phys. {\bf 10}, 469
  (1969).}}\BibitemShut {Stop}%
\bibitem [{\citenamefont {Hughes}\ and\ \citenamefont
  {Maglic}(1971)}]{hughes:1971aa}%
  \BibitemOpen
  \bibfield  {author} {\bibinfo {author} {\bibfnamefont {V.~W.}\ \bibnamefont
  {Hughes}}\ and\ \bibinfo {author} {\bibfnamefont {B.}~\bibnamefont
  {Maglic}},\ }\href@noop {} {\bibfield  {journal} {\bibinfo  {journal} {Bull.
  Am. Phys. Soc.}\ }\textbf {\bibinfo {volume} {16}},\ \bibinfo {pages} {65}
  (\bibinfo {year} {1971})}\BibitemShut {NoStop}%
\bibitem [{\citenamefont {Deutsch}(1951)}]{Deutsch:1951zza}%
  \BibitemOpen
  \bibfield  {author} {\bibinfo {author} {\bibfnamefont {M.}~\bibnamefont
  {Deutsch}},\ }\href {\doibase 10.1103/PhysRev.82.455} {\bibfield  {journal}
  {\bibinfo  {journal} {Phys. Rev.}\ }\textbf {\bibinfo {volume} {82}},\
  \bibinfo {pages} {455} (\bibinfo {year} {1951})}\BibitemShut {NoStop}%
\bibitem [{\citenamefont {Banburski}\ and\ \citenamefont
  {Schuster}(2012)}]{Banburski:2012tk}%
  \BibitemOpen
  \bibfield  {author} {\bibinfo {author} {\bibfnamefont {A.}~\bibnamefont
  {Banburski}}\ and\ \bibinfo {author} {\bibfnamefont {P.}~\bibnamefont
  {Schuster}},\ }\href {\doibase 10.1103/PhysRevD.86.093007} {\bibfield
  {journal} {\bibinfo  {journal} {Phys.~Rev.}\ }\textbf {\bibinfo {volume}
  {D86}},\ \bibinfo {pages} {093007} (\bibinfo {year} {2012})},\ \Eprint
  {http://arxiv.org/abs/1206.3961} {arXiv:1206.3961 [hep-ph]} \BibitemShut
  {NoStop}%
\bibitem [{\citenamefont {Eidelman}(2017)}]{simonPriv}%
  \BibitemOpen
  \bibfield  {author} {\bibinfo {author} {\bibfnamefont {S.}~\bibnamefont
  {Eidelman}},\ }\href@noop {} {} (\bibinfo {year} {2017}),\ \bibinfo {note}
  {private communication.}\BibitemShut {Stop}%
\bibitem [{\citenamefont {Sakimoto}(2015)}]{Sakimoto:2015vrf}%
  \BibitemOpen
  \bibfield  {author} {\bibinfo {author} {\bibfnamefont {K.}~\bibnamefont
  {Sakimoto}},\ }\href {\doibase 10.1140/epjd/e2015-60427-6} {\bibfield
  {journal} {\bibinfo  {journal} {Eur. Phys. J.}\ }\textbf {\bibinfo {volume}
  {D69}},\ \bibinfo {pages} {276} (\bibinfo {year} {2015})}\BibitemShut
  {NoStop}%
\bibitem [{\citenamefont {Brodsky}\ and\ \citenamefont
  {Lebed}(2009)}]{Brodsky:2009gx}%
  \BibitemOpen
  \bibfield  {author} {\bibinfo {author} {\bibfnamefont {S.~J.}\ \bibnamefont
  {Brodsky}}\ and\ \bibinfo {author} {\bibfnamefont {R.~F.}\ \bibnamefont
  {Lebed}},\ }\href {\doibase 10.1103/PhysRevLett.102.213401} {\bibfield
  {journal} {\bibinfo  {journal} {Phys. Rev. Lett.}\ }\textbf {\bibinfo
  {volume} {102}},\ \bibinfo {pages} {213401} (\bibinfo {year} {2009})},\
  \Eprint {http://arxiv.org/abs/0904.2225} {arXiv:0904.2225 [hep-ph]}
  \BibitemShut {NoStop}%
\bibitem [{\citenamefont {Ginzburg}\ \emph {et~al.}(1998)\citenamefont
  {Ginzburg}, \citenamefont {Jentschura}, \citenamefont {Karshenboim},
  \citenamefont {Krauss}, \citenamefont {Serbo},\ and\ \citenamefont
  {Soff}}]{Ginzburg:1998df}%
  \BibitemOpen
  \bibfield  {author} {\bibinfo {author} {\bibfnamefont {I.~F.}\ \bibnamefont
  {Ginzburg}}, \bibinfo {author} {\bibfnamefont {U.~D.}\ \bibnamefont
  {Jentschura}}, \bibinfo {author} {\bibfnamefont {S.~G.}\ \bibnamefont
  {Karshenboim}}, \bibinfo {author} {\bibfnamefont {F.}~\bibnamefont {Krauss}},
  \bibinfo {author} {\bibfnamefont {V.~G.}\ \bibnamefont {Serbo}},\ and\
  \bibinfo {author} {\bibfnamefont {G.}~\bibnamefont {Soff}},\ }\href@noop {}
  {\bibfield  {journal} {\bibinfo  {journal} {Phys. Rev.}\ }\textbf {\bibinfo
  {volume} {C58}},\ \bibinfo {pages} {3565} (\bibinfo {year} {1998})},\ \Eprint
  {http://arxiv.org/abs/hep-ph/9805375} {hep-ph/9805375} \BibitemShut {NoStop}%
\bibitem [{\citenamefont {Nemenov}(1972)}]{Nemenov:1972ph}%
  \BibitemOpen
  \bibfield  {author} {\bibinfo {author} {\bibfnamefont {L.~L.}\ \bibnamefont
  {Nemenov}},\ }\href@noop {} {\bibfield  {journal} {\bibinfo  {journal} {Yad.
  Fiz.}\ }\textbf {\bibinfo {volume} {15}},\ \bibinfo {pages} {1047} (\bibinfo
  {year} {1972})},\ \bibinfo {note} {{Sov. J. Nucl. Phys. {\bf 15}, 582
  (1972).}}\BibitemShut {Stop}%
\bibitem [{\citenamefont {Ji}\ and\ \citenamefont {Lamm}(2016)}]{Ji:2016fat}%
  \BibitemOpen
  \bibfield  {author} {\bibinfo {author} {\bibfnamefont {Y.}~\bibnamefont
  {Ji}}\ and\ \bibinfo {author} {\bibfnamefont {H.}~\bibnamefont {Lamm}},\
  }\href {\doibase 10.1103/PhysRevA.94.032507} {\bibfield  {journal} {\bibinfo
  {journal} {Phys. Rev.}\ }\textbf {\bibinfo {volume} {A94}},\ \bibinfo {pages}
  {032507} (\bibinfo {year} {2016})},\ \Eprint
  {http://arxiv.org/abs/1607.07059} {arXiv:1607.07059 [physics.atom-ph]}
  \BibitemShut {NoStop}%
\bibitem [{\citenamefont {Lamm}(2015)}]{Lamm:2015fia}%
  \BibitemOpen
  \bibfield  {author} {\bibinfo {author} {\bibfnamefont {H.}~\bibnamefont
  {Lamm}},\ }in  B.~Fleming and W.~Haxton (eds.),\ \href
  {http://inspirehep.net/record/1395477/files/arXiv:1509.09306.pdf}  {\emph
  {\bibinfo {booktitle} {{P}roceedings, 12th Conference on the Intersections of
  Particle and Nuclear Physics}}}\ (\bibinfo
  {year} { (CIPANP 2015): Vail, Colorado, USA, 2015}),  http://www.slac.stanford.edu/econf/C150519,\ \Eprint {http://arxiv.org/abs/1509.09306} {arXiv:1509.09306
  [hep-ph]} \BibitemShut {NoStop}%
\bibitem [{\citenamefont {Jentschura}\ \emph {et~al.}(1997)\citenamefont
  {Jentschura}, \citenamefont {Soff}, \citenamefont {Ivanov},\ and\
  \citenamefont {Karshenboim}}]{Jentschura:1997tv}%
  \BibitemOpen
  \bibfield  {author} {\bibinfo {author} {\bibfnamefont {U.~D.}\ \bibnamefont
  {Jentschura}}, \bibinfo {author} {\bibfnamefont {G.}~\bibnamefont {Soff}},
  \bibinfo {author} {\bibfnamefont {V.~G.}\ \bibnamefont {Ivanov}},\ and\
  \bibinfo {author} {\bibfnamefont {S.~G.}\ \bibnamefont {Karshenboim}},\
  }\href {\doibase 10.1103/PhysRevA.56.4483} {\bibfield  {journal} {\bibinfo
  {journal} {Phys. Rev.}\ }\textbf {\bibinfo {volume} {A56}},\ \bibinfo {pages}
  {4483} (\bibinfo {year} {1997})},\ \Eprint
  {http://arxiv.org/abs/physics/9706026} {arXiv:physics/9706026 [physics]}
  \BibitemShut {NoStop}%
\bibitem [{\citenamefont {Karshenboim}\ \emph
  {et~al.}(1998{\natexlab{a}})\citenamefont {Karshenboim}, \citenamefont
  {Jentschura}, \citenamefont {Ivanov},\ and\ \citenamefont
  {Soff}}]{Karshenboim:1998we}%
  \BibitemOpen
  \bibfield  {author} {\bibinfo {author} {\bibfnamefont {S.~G.}\ \bibnamefont
  {Karshenboim}}, \bibinfo {author} {\bibfnamefont {U.~D.}\ \bibnamefont
  {Jentschura}}, \bibinfo {author} {\bibfnamefont {V.~G.}\ \bibnamefont
  {Ivanov}},\ and\ \bibinfo {author} {\bibfnamefont {G.}~\bibnamefont
  {Soff}},\ }\href {\doibase 10.1016/S0370-2693(98)00206-8} {\bibfield
  {journal} {\bibinfo  {journal} {Phys. Lett.}\ }\textbf {\bibinfo {volume}
  {B424}},\ \bibinfo {pages} {397} (\bibinfo {year} {1998}{\natexlab{a}})},\
  \Eprint {http://arxiv.org/abs/hep-ph/9706401} {arXiv:hep-ph/9706401 [hep-ph]}
  \BibitemShut {NoStop}%
\bibitem [{\citenamefont {Karshenboim}\ \emph
  {et~al.}(1998{\natexlab{b}})\citenamefont {Karshenboim}, \citenamefont
  {Jentschura}, \citenamefont {Ivanov},\ and\ \citenamefont
  {Soff}}]{karshenboim:1998aa}%
  \BibitemOpen
  \bibfield  {author} {\bibinfo {author} {\bibfnamefont {S.~G.}\ \bibnamefont
  {Karshenboim}}, \bibinfo {author} {\bibfnamefont {U.~D.}~\bibnamefont
  {Jentschura}}, \bibinfo {author} {\bibfnamefont {V.~G.}\ \bibnamefont
  {Ivanov}},\ and\ \bibinfo {author} {\bibfnamefont {G.}~\bibnamefont
  {Soff}},\ }\href@noop {} {\bibfield  {journal} {\bibinfo  {journal} {Eur.~Phys.~J.}\ }\textbf {\bibinfo {volume} {D2}},\ \bibinfo {pages}
  {209} (\bibinfo {year} {1998}{\natexlab{b}})}\BibitemShut {NoStop}%
\bibitem [{\citenamefont {Malenfant}(1987)}]{Malenfant:1987tm}%
  \BibitemOpen
  \bibfield  {author} {\bibinfo {author} {\bibfnamefont {J.}~\bibnamefont
  {Malenfant}},\ }\href {\doibase 10.1103/PhysRevD.36.863} {\bibfield
  {journal} {\bibinfo  {journal} {Phys. Rev.}\ }\textbf {\bibinfo {volume}
  {D36}},\ \bibinfo {pages} {863} (\bibinfo {year} {1987})}\BibitemShut
  {NoStop}%
\bibitem [{\citenamefont {Steinberger}(1949)}]{Steinberger:1949wx}%
  \BibitemOpen
  \bibfield  {author} {\bibinfo {author} {\bibfnamefont {J.}~\bibnamefont
  {Steinberger}},\ }\href {\doibase 10.1103/PhysRev.76.1180} {\bibfield
  {journal} {\bibinfo  {journal} {Phys. Rev.}\ }\textbf {\bibinfo {volume}
  {76}},\ \bibinfo {pages} {1180} (\bibinfo {year} {1949})}\BibitemShut
  {NoStop}%
\bibitem [{\citenamefont {Blokland}\ \emph {et~al.}(2002)\citenamefont
  {Blokland}, \citenamefont {Czarnecki},\ and\ \citenamefont
  {Melnikov}}]{Blokland:2001pb}%
  \BibitemOpen
  \bibfield  {author} {\bibinfo {author} {\bibfnamefont {I.~R.}\ \bibnamefont
  {Blokland}}, \bibinfo {author} {\bibfnamefont {A.}~\bibnamefont {Czarnecki}},
 and\ \bibinfo {author} {\bibfnamefont {K.}~\bibnamefont {Melnikov}},\
  }\href {\doibase 10.1103/PhysRevLett.88.071803} {\bibfield  {journal}
  {\bibinfo  {journal} {Phys. Rev. Lett.}\ }\textbf {\bibinfo {volume} {88}},\
  \bibinfo {pages} {071803} (\bibinfo {year} {2002})},\ \Eprint
  {http://arxiv.org/abs/hep-ph/0112117} {arXiv:hep-ph/0112117} \BibitemShut
  {NoStop}%
\bibitem [{\citenamefont {Achasov}\ and\ \citenamefont
  {Kiselev}(2002{\natexlab{a}})}]{Achasov:2002bh}%
  \BibitemOpen
  \bibfield  {author} {\bibinfo {author} {\bibfnamefont {N.~N.}\ \bibnamefont
  {Achasov}}\ and\ \bibinfo {author} {\bibfnamefont {A.~V.}\ \bibnamefont
  {Kiselev}},\ }\href {\doibase 10.1103/PhysRevD.65.097302} {\bibfield
  {journal} {\bibinfo  {journal} {Phys. Rev.}\ }\textbf {\bibinfo {volume}
  {D65}},\ \bibinfo {pages} {097302} (\bibinfo {year} {2002}{\natexlab{a}})},\
  \Eprint {http://arxiv.org/abs/hep-ph/0202047} {arXiv:hep-ph/0202047 [hep-ph]}
  \BibitemShut {NoStop}%
\bibitem [{\citenamefont {Achasov}\ and\ \citenamefont
  {Kiselev}(2002{\natexlab{b}})}]{Achasov:2002xm}%
  \BibitemOpen
  \bibfield  {author} {\bibinfo {author} {\bibfnamefont {N.~N.}\ \bibnamefont
  {Achasov}}\ and\ \bibinfo {author} {\bibfnamefont {A.~V.}\ \bibnamefont
  {Kiselev}},\ }\href {\doibase 10.1134/1.1500713} {\bibfield  {journal}
  {\bibinfo  {journal} {JETP Lett.}\ }\textbf {\bibinfo {volume} {75}},\
  \bibinfo {pages} {527} (\bibinfo {year} {2002}{\natexlab{b}})},\ \bibinfo
  {note} {[Pisma Zh. Eksp. Teor. Fiz.75,643(2002)]}\BibitemShut {NoStop}%
\bibitem [{\citenamefont {Marciano}\ \emph {et~al.}(2016)\citenamefont
  {Marciano}, \citenamefont {Masiero}, \citenamefont {Paradisi},\ and\
  \citenamefont {Passera}}]{Marciano:2016yhf}%
  \BibitemOpen
  \bibfield  {author} {\bibinfo {author} {\bibfnamefont {W.~J.}\ \bibnamefont
  {Marciano}}, \bibinfo {author} {\bibfnamefont {A.}~\bibnamefont {Masiero}},
  \bibinfo {author} {\bibfnamefont {P.}~\bibnamefont {Paradisi}},\ and\
  \bibinfo {author} {\bibfnamefont {M.}~\bibnamefont {Passera}},\ }\href
  {\doibase 10.1103/PhysRevD.94.115033} {\bibfield  {journal} {\bibinfo
  {journal} {Phys. Rev.}\ }\textbf {\bibinfo {volume} {D94}},\ \bibinfo {pages}
  {115033} (\bibinfo {year} {2016})},\ \Eprint
  {http://arxiv.org/abs/1607.01022} {arXiv:1607.01022 [hep-ph]} \BibitemShut
  {NoStop}%
\bibitem{Dorokhov:2012qa}
  A.~E.~Dorokhov, A.~E.~Radzhabov and A.~S.~Zhevlakov,
  Eur.\ Phys.\ J.\ C {\bf 72}, 2227 (2012),  arXiv:1204.3729 [hep-ph].
\bibitem [{\citenamefont {Brodsky}\ and\ \citenamefont
  {de~Rafael}(1968)}]{Brodsky:1967sr}%
  \BibitemOpen
  \bibfield  {author} {\bibinfo {author} {\bibfnamefont {S.~J.}\ \bibnamefont
  {Brodsky}}\ and\ \bibinfo {author} {\bibfnamefont {E.}~\bibnamefont
  {de~Rafael}},\ }\href {\doibase 10.1103/PhysRev.168.1620} {\bibfield
  {journal} {\bibinfo  {journal} {Phys. Rev.}\ }\textbf {\bibinfo {volume}
  {168}},\ \bibinfo {pages} {1620} (\bibinfo {year} {1968})}\BibitemShut
  {NoStop}%
\end{thebibliography}
%

\end{document}